\documentclass[preprint,preprintnumbers,amsmath,amssymb]{revtex4-1}

\usepackage{graphicx}
\usepackage{comment}
\usepackage[normalem]{ulem}
\usepackage{color}
\usepackage{float}

\definecolor{darkpastelgreen}{rgb}{0.01, 0.75, 0.24}

\makeatletter
\def\blfootnote{\xdef\@thefnmark{}\@footnotetext}
\makeatother

\begin{document}

\title{Quantum oscillations in the heat capacity of Kondo insulator YbB$_{12}$ }

\author{Kuan-Wen Chen$^{1\dagger}$, Yuan Zhu$^{1\dagger}$, Danilo Ratkovski$^{2}$, Guoxin Zheng$^1$, Dechen Zhang$^{1}$, Aaron Chan$^{1}$, Kaila Jenkins$^{1}$, Joanna Blawat$^{3}$, Tomoya Asaba$^{4}$, Fumitoshi Iga$^{5}$,  C. Varma$^{6}$, Yuji Matsuda$^{4}$,  John Singleton$^{3}$, Alimamy F. Bangura$^{2}$,}
\author{Lu Li$^{1}$}

\email{luli@umich.edu}

\affiliation{
	$^1$Department of Physics, University of Michigan, Ann Arbor, MI 48109, USA\\
	$^2$National High Magnetic Field Laboratory, 1800 East Paul Dirac Drive, Tallahassee, Florida 32310-3706, USA\\
	$^3$National High Magnetic Field Laboratory, MS E536, Los Alamos National Laboratory, Los Alamos, NM 87545, USA\\
        $^4$Department of Physics, Kyoto University, Kyoto 606-8502, Japan\\
        $^5$Institute of Quantum Beam Science, Graduate School of Science and Engineering, Ibaraki University, Mito 310-8512, Japan\\
	$^6$University of California, Riverside, CA. 92521
}

\date{\today}
\begin{abstract}
We observe the magnetic quantum oscillation in the heat capacity of the Kondo insulator YbB$_{12}$. The frequency of these oscillations $F = 670$ T, aligns with findings from magnetoresistance and torque magnetometry experiments for $\mu_0 H > 35$ T in the Kondo insulating phase. Remarkably, the quantum oscillation amplitudes in the heat capacity are substantial, with $\Delta \tilde{C}/T \approx$ 0.5 $\rm{mJ}$ $\rm{mol^{-1}K^{-2}}$ at 0.8 K, accounting for 13$\%$ of the known linear heat capacity coefficient $\gamma$. Double-peak structures of quantum-oscillation amplitudes due to the distribution function of fermions were identified and used to determine the value of the effective mass from the heat capacity, which agrees well with that from torque magnetometry. These observations support charge-neutral fermions contributing to the quantum oscillations in YbB$_{12}$.

\end{abstract}

\blfootnote{$\dagger$ These authors contributed equally and shared the first authorship.\\}


\maketitle                   
\renewcommand{\thesubsection}{\Alph{subsection}}
\renewcommand{\thesubsubsection}{\textit{\alph{subsubsection}}}

Quantum oscillations have traditionally served as a direct probe of the Fermi surfaces of metals~\cite{Shoenberg}. Surprisingly, in recent years, these oscillations have also been observed in electrical insulators, where a Fermi surface was previously thought to be absent, challenging our understanding of electronic band structure. This phenomenon has been detected in insulating phases such as Kondo insulators (e.g. SmB$_6$~\cite{Li2014, Tan2015}, YbB$_{12}$~\cite{Xiang2018, Sato2019, Liu2018, Christopher2024}), insulating InAs/GaSb quantum wells~\cite{Xiao2019, Han2019}, and the kagome Mott insulator YCu$_3$(OH)$_6$Br$_2$[Br$_x$(OH)$_{1-x}$]~\cite{Zheng2023, Zheng2024}.

In the Kondo insulator YbB$_{12}$, quantum oscillations were measured by several groups independently~\cite{Xiang2018,Liu2018,Christopher2024}. The oscillations are observed in both resistivity $\rho $ (the Shubnikov-de Haas (SdH) effect) and magnetization $M$ (the de Haas-van Alphen(dHvA) effect).
The oscillations are periodic in $1/B$, and their $T$-dependence follows the Lifshitz-Kosevich(LK) formula \cite{Xiang2022}. 

Additionally, a finite linear heat capacity coefficient $\gamma$, approximately 3.8 $\rm{mJ mol^{-1}K^{-2}}$, was observed in heat-capacity measurements ~\cite{Sato2019}. This value agrees with the Fermi surface size and effective mass deduced from quantum oscillations, suggesting that the latter represents a bulk (rather than surface) phenomenon.  A finite residual $T$-linear term in the thermal conductivity is also observed  $\kappa_{xx}/T\neq 0(T \to 0)$, indicating itinerant and gapless excitations ~\cite{Sato2019}. Furthermore, a recent study observed double peak features in heat capacity of YbB$_{12}$ in applied magnetic field, providing evidence for fermionic excitations \cite{Yang2024}. However, the upper field limit was 35 T, below where quantum oscillations with $1/B$ periodicity were confirmed in magnetization and electrical resistivity \cite{Xiang2018,Liu2018, Christopher2024}. Various theoretical models have been proposed to explain types of quasiparticles that can transport heat and respond to magnetic vector potential through quantum oscillations, yet do not transport charge ~\cite{Varma2020, Varma2023, Chowdhury2018, Rao2019, Heath2020, Erten2017, Cooper2023, Erten2016, Shen2018, Harrison2018, Fuhrman2020, Knolle2017, Tada2020, Ram2017}.

However, the community has not yet reached a consensus on fundamental questions about the quantum oscillations; debates persist about whether the oscillations originate from the surface or the bulk and whether they are intrinsic or extrinsic. It is essential to obtain direct {\it thermodynamic} evidence to determine whether the above-mentioned thermal properties originate from the same quasiparticle band as the quantum oscillations.

\begin{figure*}[!htb]
\begin{center}
	\includegraphics[width= 0.9\textwidth ]{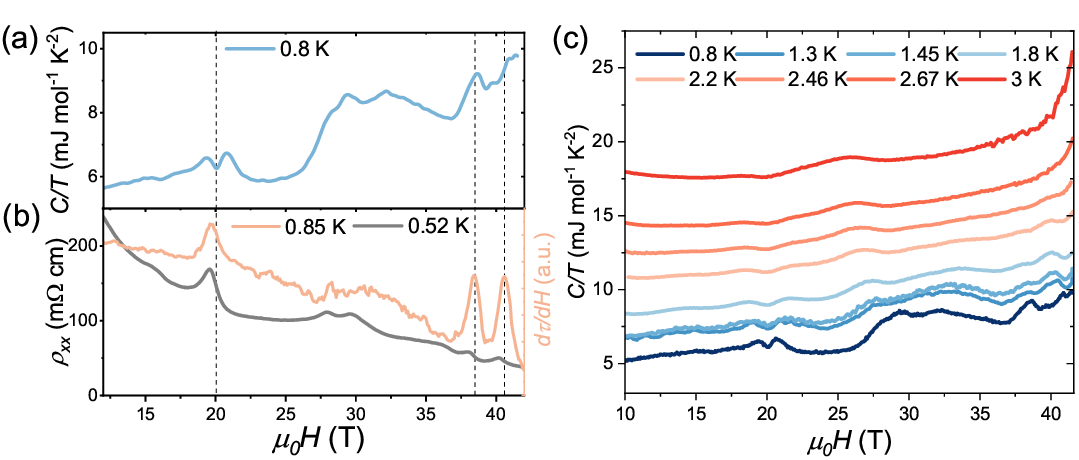}
	\caption{ (a)~Heat capacity divided by temperature $C/T$ as a function of the magnetic field at 0.8 K displays a distinctive double-peak structure at a transition at $\mu_0 H \approx 20$~T and quantum oscillations above 35~T.  (b)~Resistivity $\rho_{xx}$ and the field derivative of torque $d\tau/dH$ as function of the magnetic field at 0.52~K and 0.85~K, respectively. Dashed lines mark the location of the transition at around 20~T and quantum oscillations at about 35~T. (c)~$C/T$ as a function of the magnetic field at various temperatures.
			} 
		\label{fig:Figure1}
	\end{center}
\end{figure*}

Figure~\ref{fig:Figure1}(a) presents the variation of the heat capacity divided by temperature, $C/T$, in magnetic fields of up to 41~T obtained using a DC field magnet. The heat-capacity findings align closely with previous magnetoresistance (MR) and torque measurements, as reported in~\cite{Xiang2018} (see Fig.~\ref{fig:Figure1}(b)). Quantum oscillations with a frequency $F = 670$ T are detected in $C/T$, magnetoresistance ($\rho_{xx}$), and torque ($\tau$) measurements for $\mu_0 H > 35$ T. An electronic transition is identified near magnetic field $\mu_0 H_0 = 20$~T \cite{Xiang2022}. 
Moreover, a distinctive double-peak structure associated with both the $H_0=20$ T transition and quantum oscillations is evident in the heat capacity data.

\begin{figure*}[!htb]
	\begin{center}
		\includegraphics[width= 1\textwidth ]{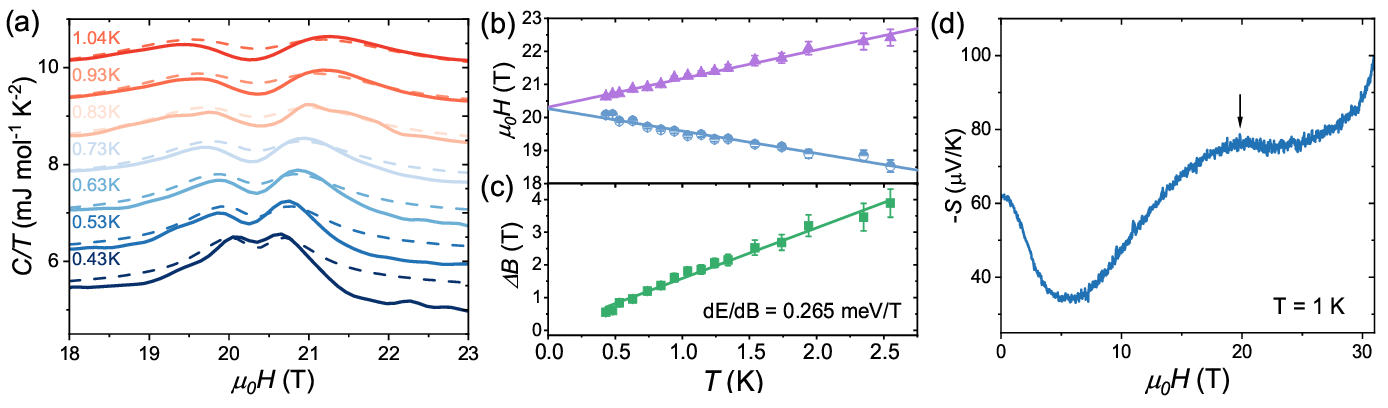}
		\caption{  (a)~The double peak structure in the heat capacity divided by temperature $C/T$  of the Lifshitz transition \cite{Xiang2022} at various temperatures. Dashed lines are fits using a single symmetric cusp-like $D(E)$ in a magnetic field, as shown in Extended Data Fig.~\ref{fig:SFigure1}. The curves are shifted vertically for clarity. (b)~The field positions of the double peaks are plotted as a function of temperature, with error bars indicating the uncertainty in defining the exact peak positions. (c)~Temperature dependence of the field separation $\Delta B$ between the double peaks. The green solid line plots $\Delta B {\rm d}E/{\rm d}B=$4.8 $k_{\rm B} T$ expected from the term $-(E-E_F)^2{\rm d}f/{\rm d}E$. (d)~Magnetic field dependence of thermopower in YbB$_{12}$ up to 31~T. The asymmetric peak splitting at around 20~T indicates a Lifshitz transition \cite{Xiang2022}. The sign of the thermopower is consistent with electron-like.
		} 
		\label{fig:Figure2}
	\end{center}
\end{figure*}

The transition at $H_0=20$ T was originally associated with the Lifshitz transition of the electronic structure in the earlier literature \cite{Xiang2022}, in which the Fermi surfaces of charge-neutral quasiparticles went through some transitions driven by the Zeeman energy, at the same time the remaining of the charged electrons scatter with the charge-neutral FS, and thus leading to similar signatures in the magnetoresistance\cite{Xiang2022}, magnetic torque\cite{Xiang2022}, and here the heat capacity. Below Fig. \ref{fig:Figure2}(c) will also show the thermopower displaying feature at the same magnetic field $H_0 = 20$ T. However, our direct measurements of the heat capacity review that the value of the heat capacity stays the same. The Hall conductivity measurement shows a consistent carrier density \cite{Xiang2022}. The earlier measurements of the heat capacity at lower fields show values of the electronic heat capacity parameter $\gamma$-term are consistent with a simple spherical model of the Fermi surface \cite{Sato2019}. This reminds us that while the Fermi Surface of copper is a large sphere with some small connecting neck orbits, the electronic heat capacity of copper is dominated by the free-electron-like spherical Fermi surface. Similarly, in the case of YbB$_{12}$,  the majority of the density of states does not change across this $H_0=20$ T transition. Possibly only some small neck orbits appear due to this 20 T transition. This transition is the origin of the dominating feature observed in the heat capacity at $H>H_0$. To be consistent with the literature, in the following text this 20 T transition will be still referred to as a Lifshitz transition.

The detection of a double-peak structure in the heat capacity at the Lifshitz transition has been observed in heavy-fermion compounds such as CeRu$_2$Si$_2$~\cite{Aoki1998} and UCoGe~\cite{Yang2023}. The existence of double-peak structures in the quantum oscillations of heat capacity has been theoretically predicted~\cite{Shao2014} and, more recently, experimentally identified in natural graphite, with a comprehensive analysis provided by Yang et al.~\cite{Yang2023}. For fermionic quasiparticles, the relationship between heat capacity over temperature and the quasiparticle energy spectrum is given by
\begin{equation}
	\frac{C}{T} = k_B^2  \int_{-\infty}^{\infty} D(E) \left( -\left(\frac{E-E_F}{k_{\rm B}T}\right)^2 \frac{\partial  f}{\partial E} \right) {\rm d}E,
 \label{HC_integral}
\end{equation}
where $f$ is Fermi-Dirac distribution function $f=1/(e^{(E-E_F)/k_{\rm B}T}+1)$ and $D(E)$ is the density of states. Unlike the kernel term $-{\rm d}f/{\rm d}E$ in electrical conductivity, the dual-peak structure originates from the kernel term $-(E-E_F)^2 {\rm d}f/{\rm d}E$ in heat capacity. Moreover, the quasiparticles probed are distributed around the chemical potential at finite temperatures, which is temperature-dependent.  The separation between the resulting maxima increases linearly with temperature as $\Delta E =$ 4.8 $k_{\rm B} T$ and disappears at zero temperature (see Extended Data Fig.~\ref{fig:SFigure1}(a)). The behavior of $C/T$ versus $\mu_0 H$ at various temperatures, depicted in Fig.~\ref{fig:Figure1}(c), comes from measurements on the same sample in NHMFL 35 T and 41 T magnets. At higher temperatures, specifically for $T > 2.67$~K, the dramatic increase in $C/T$ for $\mu_0 H > 30$~T is associated with a metal-insulator transition. The quantum oscillation amplitudes at 0.8~K are substantial, with $\Delta \tilde{C}/T \approx$ 0.5 $\rm{mJ mol^{-1}K^{-2}}$, accounting for 13$\%$ of the known linear heat capacity coefficient $\gamma \approx 3.8 ~\rm{mJ mol^{-1}K^{-2}}$ ~\cite{Sato2019}. Thus, it is plausible that the same neutral quasiparticles contribute to both the quantum oscillations and the $\gamma$ coefficient, allowing us to rule out the hypothesis that quantum oscillations arise from impurity phases \cite{Cooper2023, Fuhrman2020}.

The temperature dependence of the symmetric double-peak structure of field-induced Lifshitz transition in $C/T$ at 20~T is shown in Fig.~\ref{fig:Figure2}(a). The term $-(E-E_{\rm F})^2 {\rm d}f/{\rm d}E$ in eq.(1) is also symmetric about the Fermi energy $E_{\rm F}$. Therefore, the variation of $C/T$ with $\mu_0 H$ is effectively modeled by a symmetric density of states $D(E)$  intersecting the Fermi energy in the magnetic field, as illustrated in Extended Data Fig.~\ref{fig:SFigure1}(b)(c). The peak at lower magnetic field in $C/T$ occurs when the peak of $D(E)$ coincides with the maximum of $-(E-E_{\rm F})^2 df/dE$ at $E<E_{\rm F}$, while the peak at higher magnetic fields emerges when the peak of $D(E)$ is aligned with the other maximum of $-(E-E_{\rm F})^2 {\rm d}f/{\rm d}E$ at $E>E_{\rm F}$. The dip in $C/T$ between the double peaks happens when the maximum $D(E)$ is at the Fermi energy. The separation between the double peaks, denoted as $\Delta B$, scales linearly with $T$, and vanishes to zero at $T \rightarrow 0$ as shown in Fig.~\ref{fig:Figure2}(b) and (c). Given that the separation of the maxima in $-(E-E_{\rm F})^2 df/dE$ is $\Delta E = 4.8 k_{\rm B} T$, the rate of change between energy and magnetic field, ${\rm d}E/{\rm d}B = 0.265~\rm{meV/T}$, is calculated from the equation $\Delta B \cdot {\rm d}E/{\rm d}B = 4.8 k_{\rm B} T$. The dotted lines in Fig.~\ref{fig:Figure2}(a) are fits using ${\rm d}E/{\rm d}B = 0.265~\rm{meV/T}$ and a cusp-like $D(E)$ (Extended Data Fig.~\ref{fig:SFigure1}), exhibit a good match with observed data. We note that the observation of the double peaks at around 20~T is consistent with a recent report on heat capacity in YbB$_{12}$ by Yang et al., whose ${\rm d}E/dB = 0.258$~meV/T~\cite{Yang2024}.

\begin{figure*}[!htb]
	\begin{center}
		\includegraphics[width= 1\textwidth ]{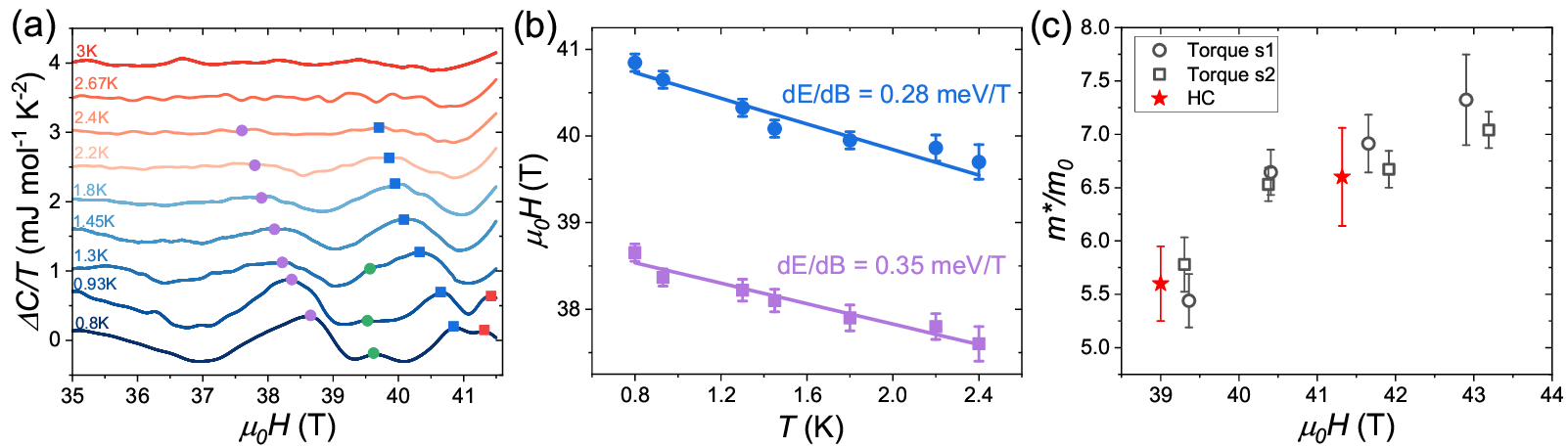}
		\caption{ (a) The oscillatory component of the heat capacity divided by temperature $\Delta C/T$ at various temperatures after subtracting the polynomial backgrounds. The curves are shifted vertically for clarity. The points mark the peak positions. (b) Magnetic field positions of the dominant peaks as a function of temperature. The lines are linear fits of the data. The error bars show the uncertainty of defining the peak positions. (c) Field dependence of the effective masses $m^*(39 ~\rm{T})=5.6 m_0$ and $m^*(41.4~\rm{T}) = 6.6 m_0$ from heat capacity compared to the effective masses obtained from the LK fits to the dHvA oscillation amplitudes in torque~\cite{Xiang2018}, where $m_0$ is the free electron mass.
		} 
		\label{fig:Figure3}
	\end{center}
\end{figure*}

The precise mechanism underlying the 20~T Lifshitz transition remains unclear at this point \cite{Xiang2022}. 
The Zeeman energy, which is directly proportional to the magnetic field can be generally expressed as $E_Z=g_J m_J \mu_{\rm B} B$, where $g_J$ and $m_J$ denote the Landé g-factor and the magnetic quantum number of the total angular momentum $J$, respectively, and $\mu_{\rm B}$ is the Bohr magneton. 

Interpreting this Lifshitz transition as driven by a single spin species crossing the Fermi energy via the Zeeman effect may be insufficient. With ${\rm d}E_Z/{\rm d}B=0.265$~meV/T, the calculated $g_Jm_J$ value is $4.58 \mu_{\rm B}$. However, the ground state arising from the crystal-field splitting of $4f$ electrons of Yb$^{3+}$ in YbB$_{12}$, characterized by $\Gamma_8$ symmetry~\cite{Nemkovski2007}, exhibits a significantly smaller $g_Jm_J$ value of 2.1 $\mu_{\rm B}$~\cite{Terashima2017}. Another contributing factor to the ${\rm d}E/{\rm d}B$ shift could be a Fermi energy displacement occurring near the Lifshitz transition. A similar double-peak structure in a Lifshitz transition was measured in the heavy fermion compound CeCu$_2$Si$_2$ and was attributed to a Fermi-energy shift~\cite{Yang2023, Matsuda2012}.

Thermoelectric effects are a powerful probe for detecting Lifshitz transitions.  We measured the Seebeck coefficient of YbB$_{12}$ in a 31~T magnet, as described in the Methods section. Lifshitz transitions are well-known for leading to a significant thermopower enhancement through the “second-and-half order” transition~\cite{Abrikosov17}. In Fig.~\ref{fig:Figure2}(d), the negative Seebeck coefficient $S$ indicates that electron-like carriers dominate, consistent with the negative Hall signal above T = 1.5 K as reported in ref.~\cite{Xiang2022}. We observed a distinct asymmetric peak near 20~T, which can be described using the expression for the Seebeck coefficient  $S = \frac{2ek_{\rm B}}{h}\int_{-\infty}^{\infty}D(E)\left(-\frac{E-E_{\rm F}}{k_{\rm B}T} \frac{\partial  f}{\partial E} \right) {\rm d}E$, where the kernel term $-(E-E_{\rm F}){\rm d}f/{\rm d}E$ introduces asymmetry (Extended Data Fig. \ref{fig:SFigure1}(d)). 

It is noted that no signature of the Lifshitz transition was observed in magnetization or using the proximity detector oscillator method (PDO)  in YbB$_{12}$ at 20 T \cite{Xiang2021}, while torque and heat capacity measurements detect features \cite{Xiang2022}.  The absence of 20 T Lifshitz transition in YbB$_{12}$ magnetization is distinct from other heavy fermion compounds that exhibit Lifshitz transitions, such as CeRu$_2$Si$_2$, where the magnetization shows a kink at the transition field, resulting in a peak in the magnetic susceptibility \cite{Flouquet2002,Boucherle2001}. In the CeRu$_2$Si$_2$ case, magnetization takes the form $M=\mu_B^2 B \int_{-\infty}^{\infty}D(E)(-\frac{\partial f}{\partial E}){\rm d}E$ when a single DOS peak moves with the magnetic field; therefore, magnetic susceptibility shows a single peak. The absence of magnetization features near 20 T thus makes the Lifshitz transition in YbB$_{12}$ unusual. One possible reason for the presence of Lifshitz transition in torque measurements near 20 T but not in $M-H$ is that magnetization $M=\chi \cdot B$ measures the diagonal component of the field-dependent susceptibility, while torque $\tau = \vec{M} \times \vec{B}$ picks up the off-diagonal component of $\chi$. The off-diagonal susceptibility could lead to a finite $\vec{M} \times \vec{B}$ term to be detected by torque at the Lifshitz transition - although the exact mechanism is still unknown - while not appearing in magnetization measurements. It also remains possible that the resolution of magnetization and PDO measurements in the pulse magnetic field were not enough to resolve the Lifshitz transition features. Magnetization measurements with high precision might be necessary to detect the transition feature near 20 T.

In Fig.~\ref{fig:Figure3}(a), $\Delta C/T$ exhibits the $F=$ 670~T quantum oscillations also seen in resistivity and torque once a polynomial background is subtracted. Two double-peak structures can be clearly observed, one centered at $\mu_0 H= 39$ T and the other at 
$\mu_0 H= 41.4$~T, corresponding to Landau indices $n = 17$ and $n = 16$ respectively. Unlike the Lifshitz transition, the double-peak structure in quantum oscillations is asymmetric. The peak at a lower magnetic field is more prominent, while the peak at a higher magnetic field is less intense. This asymmetry could be attributed to the asymmetric density of states $D(E)$ in the 3D Landau levels of quasiparticles in YbB$_{12}$. As shown in Extended Data Fig.~\ref{fig:SFigure2}, it is well known that the energy of the band electron is quantized into Landau levels in the plane perpendicular to the magnetic field. The "sawtooth" structure results from a one-dimensional density of states from the motion parallel to $\textbf{H}$ ($\text{DOS} \propto 1/\sqrt{E}$) superimposed on the $\delta$-function-like Landau levels. The quasiparticles contributing to the quantum oscillations have asymmetric 3D Landau levels similar to band electrons (Fermions). 

Fermions obey the Fermi-Dirac distribution.  Therefore, the smearing of quantum oscillation amplitude due to increasing temperature follows a specific form given by Lifshitz-Kosevich (LK) theory. Thermodynamic quantities such as magnetization and heat capacity arise from oscillations in the thermodynamic potential $\tilde{\Omega} = \tilde{\Omega}_0 R_T(T)$, where $\tilde{\Omega}_0$ is the zero temperature potential. The damping factor $R_T = X_p/\sinh(z)$ is a direct consequence of the smearing of the Fermi–Dirac distribution function near the Fermi energy, where $z = 2\pi^2 p k_{\rm B} m^* T/e \hbar B$, $m^*$ is the effective mass, $m_e$ is the free electron mass, and $p$ is an integer denoting the harmonic. Consequently, the oscillations in magnetization, $\tilde{M} = -\frac{\partial \tilde{\Omega}}{\partial B}$, are proportional to $R_T$.  The $F=$ 670 T quantum oscillation in torque $\Delta \tau$ and magnetoresistivity $\Delta \rho_{xx}$ has been shown to be well described by the $R_T$ in the LK formula by Xiang et al.~\cite{Xiang2018, Xiang2022}, consistent with Fermionic quasiparticles. Regarding the heat capacity, $\Delta \tilde{C}/T = - \frac{\partial}{\partial T} \left( \frac{\partial \tilde{\Omega}}{\partial T} \right)$, the damping factor is defined as $R_T''(z) = z (\left( \frac{2 \cosh(z)}{\sinh^2(z)} \right) - z(1 + \frac{\cosh^2(z)}{\sinh^3(z)}))$~\cite{McCombe1967,Sullivan1968}. $\Delta \tilde{C}/T$ changes sign and exhibits a ``$\pi$-phase shift" at $z=1.6$ where $R_T''(z=1.6)=0$ (Extended Data Fig.~\ref{fig:SFigure3}). This point is commonly used to determine the effective mass $m^*$~\cite{Riggs2011, Bondarenko2001}. However, such a ``$\pi$-phase shift" is not discernible in the double-peak structure. The experimental observation of a single- or double-peak structure depends on the measured range of temperatures, the effective mass, and the Landau level width.  Here, we introduce a method to determine the effective mass via the double-peak structure of the heat capacity. Since the peak at the higher magnetic field in a double-peak structure is identifiable only at temperatures below $T=1.3$ K for both Landau levels ($n = 17$ and $n = 16$), we plotted only the positions of the dominant lower field peaks against temperature in Fig.~\ref{fig:Figure3}(b), revealing a linear relationship. $\Delta E=\Delta B {\rm d}E/{\rm d}B\approx 2.4k_{\rm B} T$ should be satisfied, as shown in Extended Data Fig. \ref{fig:SFigure1}(a), where $\Delta E$ is the shift of the dominant lower field peak from zero temperature. By fitting the slope in Fig.~\ref{fig:Figure3}(b), ${\rm d}E/{\rm d}B= 0.35$~meV/T for $n = 17$ and 0.28~meV/T for $n = 1$6 were deduced. From the Landau-level energy $E=n\hbar\omega_{\rm c}$, where $\omega_{\rm c}=eB/m^*$, we are able to obtain the effective masses $m^*(n=17)= 5.6 m_0$ and $m^*(n=16)= 6.6 m_0$. These results are in good agreement with the previous dHvA torque data, as shown in  Fig.~\ref{fig:Figure3}(c).

\begin{figure*}[!htb]
	\begin{center}
		\includegraphics[width= 1\textwidth ]{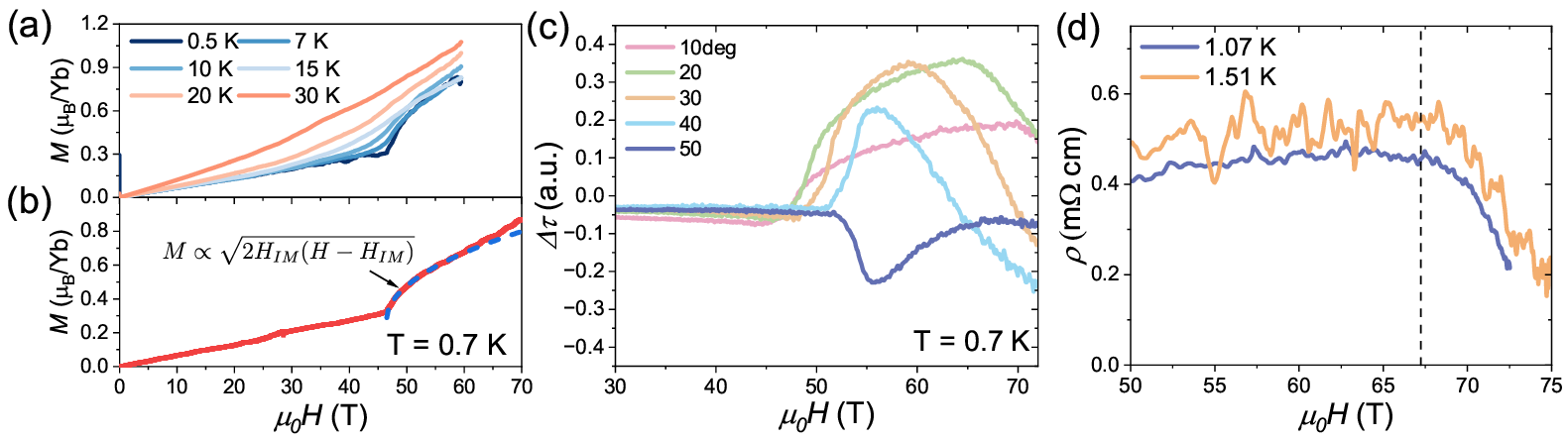}
		\caption{  (a) Magnetization measurement at various temperatures using a compensated-coil extraction magnetometer in a 60~T pulsed magnet. The field is applied along the [100] direction. (b) Magnetization measurement at 0.7~K in the 75~T Duplex magnet. The fitted curve, depicted in blue, represents $\sqrt{2H_{\rm{IM}}(H - H_{\rm{IM}})}$ where  $H_{\rm{IM}}=$46.5~T is the I-M transition field.  (c) Torque $\Delta \tau$ with a linear background subtracted at various angles from H $\parallel$ [100] (0 degrees) to H $\parallel$ [001] (90 degrees). (d) Magneto-resistivity measurement in the 75~T Duplex magnet~\cite{Xiang2021}. A downward kink is observed at 67~T.
				} 
		\label{fig:Figure4}
	\end{center}
\end{figure*}

An insulator-to-metal (I-M) transition at fields between $\mu_0 H_{\rm{IM}} \approx 45-47$~T at low temperatures is shown in Fig.~\ref{fig:Figure4}(a) and has been reported in literature~\cite{Xiang2021,Sugiyama1988,Terashima2017}. The quantum oscillations in the field-induced metallic state in YbB$_{12}$ have been observed independently by groups using proximity detector oscillator (PDO)~\cite{Xiang2021, Liu2022}, and tunnel-diode oscillator (TDO) techniques~\cite{Christopher2024}. It has been indicated that the same quasiparticle band gives rise to quantum oscillations in both insulating and metallic states. It is noted that we neglected the Zeeman energy (i.e. $E_Z=g_J m_J \mu_{\rm B} B$ = 0) in ${\rm d}E/{\rm d}B$ in heat capacity for the insulating state and derived the consistent effective masses with the ones in torque measurements by LK formula. Furthermore, the spin-splitting of Landau level is not observed in heat capacity in the insulating state, which aligns with previous quantum oscillation observations in torque $\tau$ and resistivity $\rho_{xx}$ indicating the absence of spin-splitting effects in the Kondo insulating state which was predicted by spin-less (as well as charge-less) fermions in ref \cite{Varma2020, Varma2023}.  In contrast, spin-splitting emerges in the Kondo metallic states, characterized by a $g$-factor of 0.084~\cite{Xiang2021}. In this Kondo metallic state, neutral particles start to decay into electrons and holes once the field exceeds $\mu_0 H_{\rm{IM}}$. This process is accompanied by the appearance of Zeeman splitting, which contributes to the magnetization. This contribution can be expressed as magnetization $M \propto \sqrt{2H_{\rm{IM}}(H - H_{\rm{IM}})}$ in ref \cite{Varma2023}. In Fig.~\ref{fig:Figure4}(b), a sharp increase of the magnetization shows up right after the I-M transition.  Beyond 65~T, the magnetization deviates from the $\sqrt{2H_{\rm{IM}}(H - H_{\rm{IM}})}$  and has a paramagnetic linear behavior instead.  A more pronounced feature is shown in the torque measurements, which are also sensitive to the anisotropy of magnetic susceptibility. As shown in Fig.~\ref{fig:Figure4}(c), there is a significant increase in torque at the I-M transition and a subsequent decrease at the higher field (65 - 70~T for angles close to H $\parallel$ [100] ). The magnitude of $M$ does not exhibit a sharp change, indicating that the total density of states in YbB$_{12}$ does not undergo a dramatic change. The linear field dependence of $M$ and the notable decrease in $\Delta \tau$ imply that $M$ rotates and tends to align with the direction of the magnetic field. According to the two-fluid scenario in ref.~\cite{Xiang2022}, in the Kondo metal state, fermions scatter between charge-neutral Landau levels and heavy-fermion metal bands, leading to the SdH effect in $\rho_{xx}$. 
The emergence of the new state implies the possible end of the neutral fermions, where the elimination of the density of states (DOS) for these fermions leads to reduced scattering, thereby leading to a decrease in  $\rho_{xx}$ as shown in Fig.~\ref{fig:Figure4}(d).  

In conclusion,  quantum oscillations are observed in the heat capacity in insulating YbB$_{12}$. The frequency $F = 670$~T can be derived from the inverse magnetic field of the double peak centers.  The effective masses are derived using a method based on the linear relation between double-peak separation and temperature, which are consistent with the magnetoresistance and torque via the LK formula. The large quantum oscillation amplitude (13 $\%$ of the known linear heat capacity coefficient $\gamma$) is evidence for an intrinsic bulk signal,  strongly supporting itinerant, charge-neutral excitations in YbB$_{12}$. The Liftshiz transition around $B$ = 20~T is confirmed by heat-capacity and thermopower measurements. In the extreme magnetic field beyond $\mu_0 H$ = 65 T, the magnetization $M$ deviates from the $\sqrt{2H_{\rm{IM}}(H - H_{\rm{IM}})}$ dependence and has a paramagnetic linear behavior; a pronounced feature is also shown in torque and resistivity, indicating the emergence of a new state where the magnetization $M$ rotates and tends to align with the direction of the applied magnetic field. 
The results presented herein help to narrow down existing theoretical models and stimulate future experimental and theoretical studies.

{\bf Methods\\}	
The heat capacity measurements on single crystals of YbB$_{12}$ were performed in 35/41~T resistive magnets in NHMFL, Tallahassee using high-resolution differential nanocalorimeters and an AC method \cite{Tagliati2012}. The nanocalorimeter is composed of two cells for sample and reference measurements, each consisting of electrically insulated layers of thermometer, AC-powered heater, and offset heater. The sample is in direct contact with the thermometer. The offset heater is driven by a DC current to control the temperature of the sample. The sample temperature oscillates, driven by the resistive heater with AC current of a frequency of $\omega/2$ and power $P(t)=P_{0}(1+\textrm{sin}(\omega t))$. The resulting temperature oscillation is $T_{\rm ac}(t)$, and its steady-state amplitude of the oscillating sample temperature $T_{\rm ac,0}$ measured by the thermometer can be expressed as \cite{Tagliati2012}:
\begin{equation}
	T_{\rm ac,0}=\frac{P_0}{\sqrt{(\omega C)^{2}+K^2}}
\end{equation}
and the phase difference $\phi$ between $T_{\rm ac}(t)$ and $P(t)$ is:
\begin{equation}
	{\rm tan}\phi=\frac{\omega C}{K}
\end{equation}
where $K$ is the thermal conductance between the sample and cell platform, and $C$ is the sample heat capacity. The heat capacity of the sample can then be calculated from $P_{0}$ and $T_{\rm ac,0}$ as:
\begin{equation}
	C=\frac{P_{0}}{\omega T_{\rm ac,0}} \textrm{sin}(\phi)
\end{equation}
Magnetization measurements were performed using a compensated-coil extraction magnetometer in pulsed-field magnets, giving fields up to 60/73~T in NHMFL, Los Alamos.
Magnetic torque measurements were performed using a membrane-type surface-stress sensor in the 73~T pulsed field magnet.
The electrical resistivity measurement was performed in the 75~T pulse field magnet using the technique described in Ref.\cite{Xiang2021}. The current is switched on only in the Kondo metal state above 47~T. The data taken in the up-sweep are shown in the main text since the down-sweep process suffers from Joule heating of the current. 

The thermopower measurement was done in the 31~T resistive magnet at Tallahassee, Florida. A resistive heater was attached to one end of the sample to add heat gradient, and two Lakeshore Cryotronics RX102A thermometers were thermally attached to the sample to measure the longitudinal temperature gradient. The other end of the sample was thermally anchored to a copper heat bath. Two gold wires were electrically attached to the sample to pick up the Seebeck voltage. The magnetic field was applied along the c-axis of the sample. In our measurement of the Seebeck coefficient, hole-like carrier is defined as positive.

{\bf Acknowledgement}
The work at the University of Michigan is primarily supported by the National Science Foundation under Award No.DMR-2317618 (thermodynamic measurements and thermoelectric property measurements) to Kuan-Wen Chen, Dechen Zhang, Guoxin Zheng, Aaron Chan, Yuan Zhu, Kaila Jenkins, and Lu Li.  The magnetization measurements at the University of Michigan are supported by the Department of Energy under Award No. DE-SC0020184. A portion of this work was performed at the National High Magnetic Field Laboratory (NHMFL), which is supported by National Science Foundation Cooperative Agreement Nos. DMR-1644779 and DMR-2128556 and the Department of Energy (DOE). J.S. acknowledges support from the DOE BES program “Science at 100 T,” which permitted the design and construction of much of the specialized equipment used in the high-field studies. The experiment in NHMFL is funded in part by a QuantEmX grant from ICAM and the Gordon and Betty Moore Foundation through Grant No. GBMF5305 to Kuan-Wen Chen, Dechen Zhang, Guoxin Zheng, Aaron Chan, Yuan Zhu, and Kaila Jenkins. This work at Kyoto University was supported in part by the Japan Society for the Promotion of Science (JSPS) KAKENHI Grant Number 23H00089 and the Japan Science and Technology Agency (JST) CREST Grant Number JPMJCR19T5 awarded to Yuji Matsuda.

\newpage

\setcounter{figure}{0} 
\renewcommand{\figurename}{{\bf Extended Data Fig.}}

\begin{figure}[ht]
	\centering
	\includegraphics[width=\linewidth]{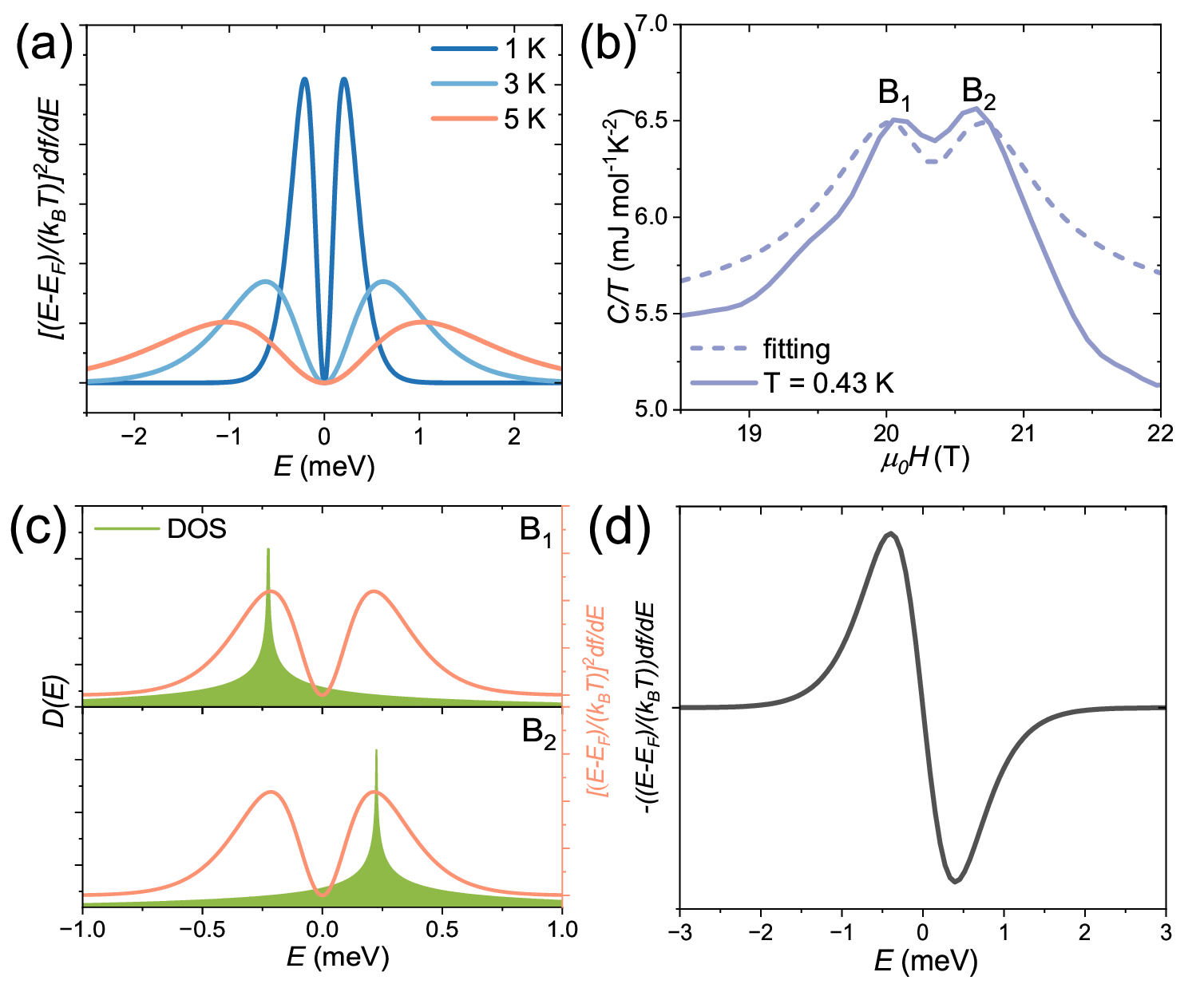}
	\caption{(a) A plot of the integral term $((E-E_F)/(k_B T))^2 df/dE$ versus energy at different temperatures. The distance between the peaks $\Delta B$ increases with temperature as $\Delta E=\Delta B dE/dB=4.8k_B T$. (b) The double peak structure of heat capacity divided by temperature $C/T$ at the Lifshitz transition at 0.43 K. The solid line shows the experimental data, and the dashed line represents the result of the fitting using Eqn. \ref{HC_integral}, assuming symmetric $D(E)$ as shown in (c). $B_1$ and $B_2$ mark the peak positions. (c) An illustration of the origin of the double peak structure in heat capacity. The green peak represents a symmetric density of state $D(E)\propto 1/\sqrt{E}$ and the orange curve shows the term $((E-E_F)/(k_B T))^2 df/dE$. The position of the DOS peak shifts with the applied magnetic field, and the convolution of the two terms produces double peaks observed in $C/T$ at the Lifshitz transition. At $B_1$ and $B_2$ the peak of $D(E)$ aligns with the maxima in the integral term. (d) An illustration of the asymmetric integral term $-(E-E_{\rm F})/(k_{\rm B}T) df/dE $ versus energy for thermopower. }
	\label{fig:SFigure1}
\end{figure}

\newpage 

\begin{figure}[ht]
	\centering
	\includegraphics[width=\linewidth]{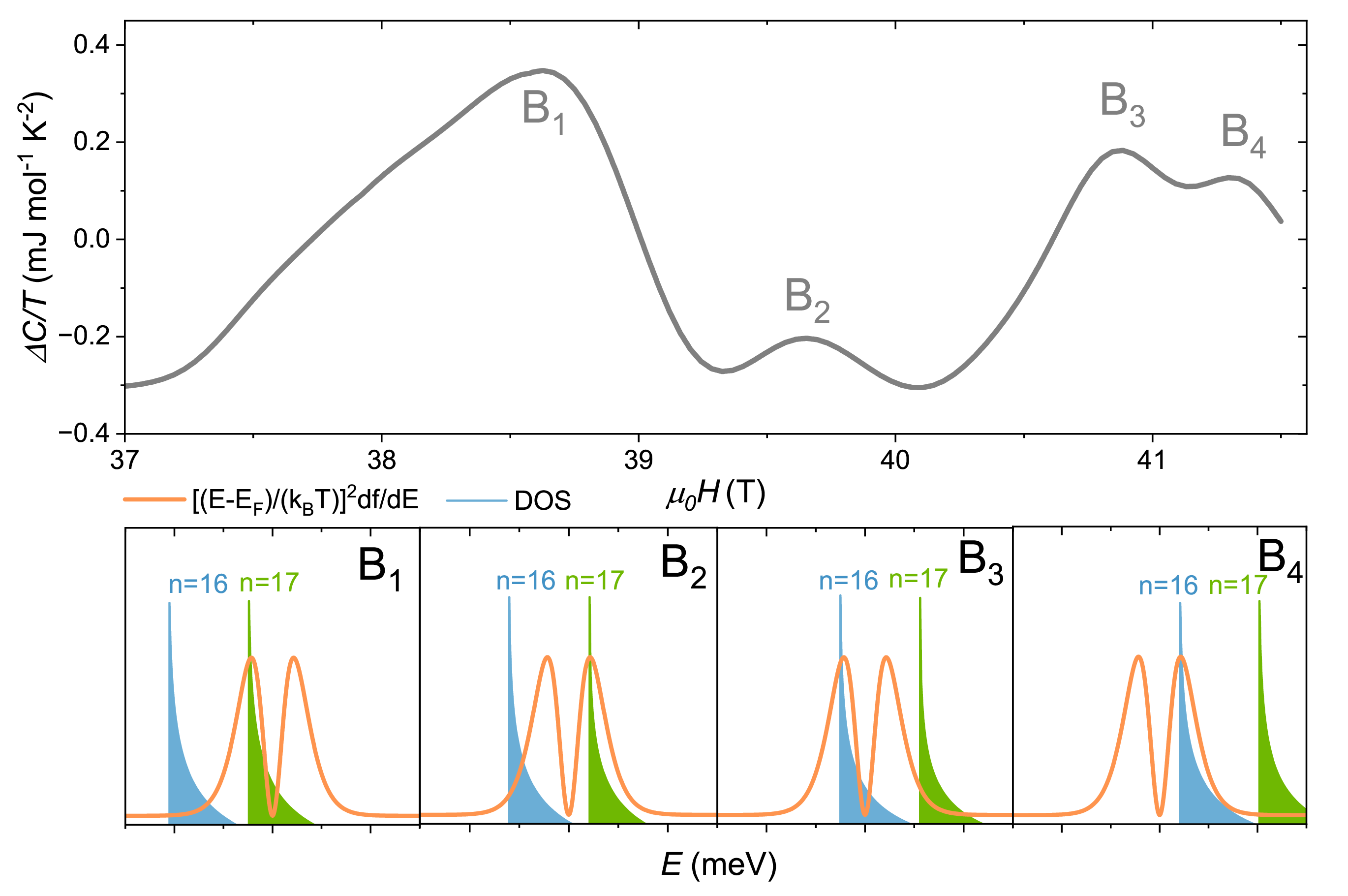}
	\caption{Top panel shows the oscillatory component of heat capacity $\Delta C/T$ at 0.8 K as a function of the magnetic field after the background subtraction. The peak positions are marked with $B_{1-4}$. The bottom panel plots the Landau level positions for $n=16$ (blue) and $n=17$ (green) with respect to the term $((E-E_F)/(k_B T))^2 df/dE$ at $B=B_{1-4}$, corresponding to the magnetic field value when one of the landau level peaks aligns with one maximum of the double peaks. }
	\label{fig:SFigure2}
\end{figure}

\begin{figure}[ht]
	\centering
	\includegraphics[width=0.5\linewidth]{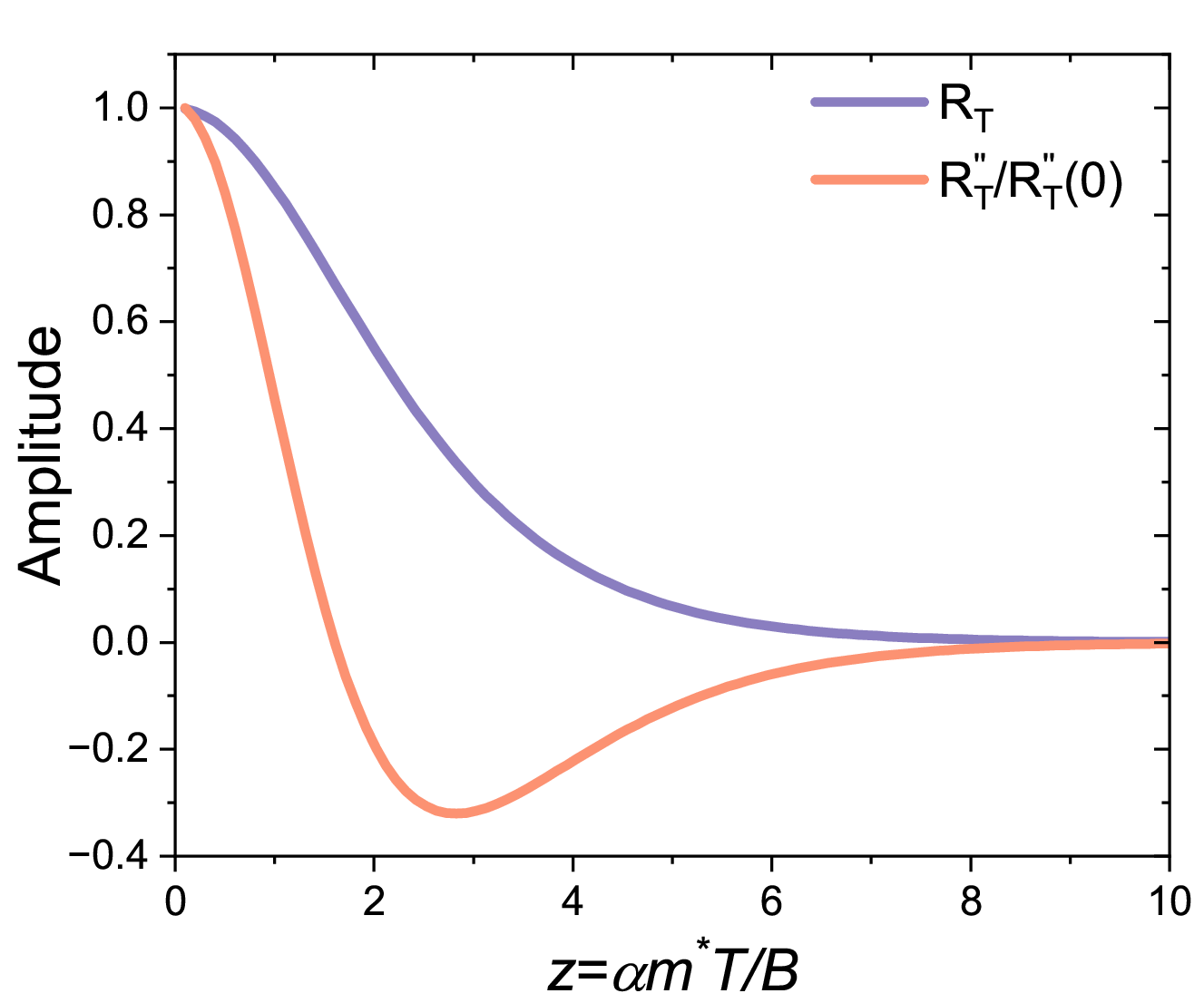}
	\caption{The normalized amplitude of the LK formula damping term $R_{T}$ and second derivative $R_{T}''$ as a function of temperature. The term $R_{T}''(z)$ where $z=\alpha m^{*} T/B$ and $\alpha=2 \pi^2 p k_B/e \hbar$ shows a sign change at $z=1.6$. }
	\label{fig:SFigure3}
\end{figure}



{\bf References}
\begin{thebibliography}{999}
\bibitem{Shoenberg}
D. Shoenberg, Magnetic Oscillations in Metals, Cambridge Monographs on Physics. Cambridge University Press (1984).
	
\bibitem{Li2014}	
G. Li et al., Two-dimensional Fermi surfaces in Kondo insulator SmB$_6$. Science, 346, 1208 (2014).
	
\bibitem{Tan2015}
B. S. Tan et al.,  Unconventional Fermi surface in an insulating state, Science {\bf 349}, 287 (2015)
	
	
\bibitem{Xiang2018}
Z. Xiang et al., Quantum oscillations of electrical resistivity in an insulator. Science {\bf 362}, 65 (2018).


\bibitem{Liu2018}
H. Liu, M. Hartstein, G. J. Wallace, A. J. Davies, M. C. Hatnean, M. D. Johannes, et al. Fermi surfaces in Kondo insulators, Journal of Physics: Condensed Matter {\bf 30}, 16LT01 (2018).


\bibitem{Sato2019}
Y. Sato et al., Unconventional thermal metallic state of charge neutral fermions in an insulator, Nat. Phys. {\bf 15}, 954-959 (2019)

\bibitem{Christopher2024}
A. Mizzi et al., The reverse quantum limit and its implications for unconventional quantum oscillations in YbB$_{12}$ Nat. Comm. {\bf 15}, 1607 (2024).


\bibitem{Xiang2022}
Z. Xiang et al., Hall Anomaly, Quantum Oscillations and Possible Lifshitz Transitions in Kondo Insulator YbB$_{12}$: Evidence for Unconventional Charge Transport. Phys. Rev. X {\bf 12}, 021050 (2022).


\bibitem{Xiao2019}
D. Xiao et al., Anomalous quantum oscillations of interacting
electron-hole gases in inverted type-II InAs/GaSb quantum wells, Phys. Rev. Lett. {\bf 122}, 186802(2019)

\bibitem{Han2019}
Z. Han, T. Li, L. Zhang, G. Sullivan and R.-R. Du, Anomalous conductance oscillations in the hybridization gap of InAs/GaSb quantum wells, Phys. Rev. Lett. {\bf 123}, 126803 (2019)


\bibitem{Zheng2023}
G. Zheng et al., Unconventional Magnetic Oscillations in Kagome Mott Insulators, arXiv: 2310.07989 (2023)

\bibitem{Zheng2024}
G. Zheng et al., Thermodynamic evidence of fermionic behavior in the vicinity of one-ninth plateau in a kagome antiferromagnet,	arXiv:2409.05600 (2024)

\bibitem{Yang2024}
Yang, Z., Marcenat, C., Kim, S. et al. Evidence for large thermodynamic signatures of in-gap fermionic quasiparticle states in a Kondo insulator. Nat Commun 15, 7801 (2024).

\bibitem{Varma2020}
C. M. Varma, Majoranas in mixed-valence insulators, Phys. Rev. B 102, 155145 (2020)

\bibitem{Varma2023}
C. M. Varma, Dark Fermions in Mixed Valence Insulators arXiv:2306.14124 (2023).

\bibitem{Chowdhury2018}
D. Chowdhury, I. Sodemann, T. Senthil, Mixed-valence insulators with neutral Fermi surfaces, Nature Commun. 9, 1766 (2018) 

\bibitem{Rao2019}
P. Rao and I. Sodemann, Cyclotron resonance inside the Mott gap: A fingerprint of emergent neutral fermions, Phys. Rev. B 100, 155150 (2019)

\bibitem{Heath2020}
J. T. Heath, K. S. Bedell,  Universal signatures of Majorana-like quasiparticles in strongly correlated Landau–Fermi liquids, J. Phys.: Condens. Matter 32 485602 (2020)

\bibitem{Erten2017}
O. Erten et al., Skyrme Insulators: Insulators at the Brink of Superconductivity, Phys. Rev. Lett. 119, 057603 (2017)

\bibitem{Cooper2023}
N. R. Cooper, J. Kelsall, Quantum oscillations in an impurity-band Anderson insulator, SciPost Phys. 15, 118 (2023) 

\bibitem{Erten2016}
O. Erten, P. Ghaemi, and P. Coleman, Kondo Breakdown and Quantum Oscillations in SmB6, Phys. Rev. Lett. 116, 046403 (2016)

\bibitem{Shen2018}
H. Shen and L. Fu, Quantum Oscillation from In-Gap States and a Non-Hermitian Landau Level Problem, Phys. Rev. Lett. 121, 026403 (2018)

\bibitem{Harrison2018}
N. Harrison, Highly Asymmetric Nodal Semimetal in Bulk SmB6, Phys. Rev. Lett. 121, 026602 (2018)

\bibitem{Knolle2017}
J. Knolle and N. R. Cooper, Excitons in topological Kondo insulators: Theory of thermodynamic and transport anomalies in SmB6, Phys. Rev. Lett. 118, 096604 (2017)

\bibitem{Fuhrman2020}
W. T. Fuhrman and P. Nikolić, Magnetic impurities in Kondo insulators: An application to samarium hexaboride, Phys. Rev. B 101, 245118 (2020)

\bibitem{Tada2020}
Y. Tada, Cyclotron resonance in Kondo insulator, Phys. Rev. Research 2, 023194 (2020)

\bibitem{Ram2017}
P. Ram and B. Kumar, Theory of quantum oscillations of magnetization in Kondo insulators, Phys. Rev. B 96, 075115 (2017)


\bibitem{Xiang2021}
Z. Xiang et al., Unusual high-field metal in a Kondo insulator, Nat. Phys. {\bf 17},788-73 (2021).

\bibitem{Flouquet2002}
J. Flouquet, et al. Itinerant metamagnetism of CeRu$_2$Si$_2$: bringing out the dead. Comparison with the new Sr$_3$Ru$_2$O$_7$ case. Physica B: Condensed Matter 319.1-4: 251-261 (2002).

\bibitem{Boucherle2001}
J. X. Boucherle, et al. Magnetic form factor in CeRu$_2$Si$_2$ on crossing its metamagnetic transition. Journal of Physics: Condensed Matter 13.48: 10901 (2001).




\bibitem{Liu2022}
H. Liu, Fermi surface in magnetic field-induced metallic YbB$_{12}$, npj Quantum Materials 7, 12 (2022)



\bibitem{Aoki1998}
Y. Aoki et al., Thermal properties of metamagnetic transition in heavy fermion systems, J. Magn. Magn. Mater. {\bf 177}, 271 (1998).

\bibitem{Yang2023}
Z. Yang et al., Unveiling the double-peak structure of
quantum oscillations in the specific heat, Nat. Comm. {\bf 14}, 7006 (2023).


\bibitem{Nemkovski2007}
K. S. Nemkovski et al., Polarized-Neutron Study of Spin Dynamics in the Kondo Insulator YbB$_{12}$. Phys. Rev. Lett. {\bf 99}, 137204 (2007)

\bibitem{Terashima2017}
T. T. Terashima et al., Magnetization process of the Kondo insulator YbB$_{12}$ in 
ultrahigh magnetic fields. J. Phys. Soc. Jpn {\bf 86}, 054710 (2017)

\bibitem{Shao2014}
Z. Shao et al., Acta Phys. Sin. {\bf 63}, 240502 (2014).

\bibitem{Matsuda2012}
Y. H. Matsuda et al. Suppression off-electron itinerancy in CeRu2Si2 by a strong magnetic field. Phys. Rev. B {\bf 86}, 041109 (2012).


\bibitem{Abrikosov17}
A. A. Abrikosov, Fundamentals of the Theory of Metals: Courier Dover Publications, (2017).

\bibitem{Behnia2015}
K. Behnia. Fundamentals of thermoelectricity. OUP Oxford, (2015).

\bibitem{Pfau2017}
H. Pfau et al., Cascade of Magnetic-Field-Induced Lifshitz Transitions in the Ferromagnetic Kondo Lattice Material 
YbNi$_4$P$_2$, Phys. Rev. Lett. {\bf 119}, 126402 (2017).

\bibitem{McCombe1967}
B. McCombe and G. Seidel, Phys. Rev. {\bf 155}, 633 (1967).

\bibitem{Sullivan1968}
P. F. Sullivan and G. Seidel, Phys. Rev. {\bf 173}, 679 (1968).



\bibitem{Riggs2011}
S. C. Riggs et al., Heat capacity through the magnetic-field-induced resistive transition in an underdoped high-temperature superconductor, Nat. Phys. {\bf 7}, 332 (2011).

\bibitem{Bondarenko2001}
V. Bondarenko et al., First observations of the heat capacity quantum oscillations in the organic superconductor (BEDT-TTF)$_2$Cu(NCS)$_2$, Synth. Met. {\bf 120}, 1039 (2001).



\bibitem{Sugiyama1988}
K. Sugiyama et al., Field-induced metallic state in YbB$_{12}$ under high magnetic field. J. Phys. Soc. Jpn {\bf 57}, 3946–3953 (1988).




\bibitem{Tagliati2012}
S. Tagliati, V. M. Krasnov, and A. Rydh,  Rev. Sci. Instrum. 83, 055107 (2012).

\end{thebibliography}
\end{document}